\documentclass[%
 aip,
 jcp,%
 amsmath,amssymb,
 floatfix,
 reprint,%
]{revtex4-1}

\usepackage{graphicx}%
\usepackage{dcolumn}%
\usepackage{bm}%
\usepackage{physics}
\usepackage{comment}

\newcommand{\br}{\mathbf{r}}
\newcommand{\bu}{\mathbf{u}}

\newcommand{\bx}{\mathbf{x}}
\newcommand{\bs}{\mathbf{s}}

\newcommand{\CX}{\mathcal{X}}

\newcommand{\CA}{\mathcal{A}}

\newcommand{\RR}{\mathbb{R}}

\begin{document}

\title{Perspective: 
Unsupervised machine learning in atomistic simulations, between predictions and understanding}

\author{Michele Ceriotti}
 \affiliation{Laboratory of Computational Science and Modeling, Institute des Materiaux, \'Ecole Polytechnique F\'ed\'erale de Lausanne, 1015 Lausanne, Switzerland}
 \email{michele.ceriotti@epfl.ch}
 
\date{\today}%

\begin{abstract}
Automated analyses of the outcome of a simulation have been an important part of atomistic modeling since the early days, addressing the need of linking the behavior of individual atoms and the collective properties that are usually the final quantity of interest.
Methods such as clustering and dimensionality reduction have been used to provide a simplified, coarse-grained representation of the structure and dynamics of complex systems, from proteins to nanoparticles. 
In recent years, the rise of machine learning has led to an even more widespread use of these algorithms in atomistic modeling, and to consider different classification and inference techniques as part of a coherent toolbox of data-driven approaches.

This perspective briefly reviews some of the unsupervised machine-learning methods -- that are geared towards classification and coarse-graining of molecular simulations --  seen in relation to the fundamental mathematical concepts that underlie all machine-learning techniques.
It discusses the importance of using concise yet complete representations of atomic structures as the starting point of the analyses, and highlights the risk of introducing preconceived biases when using machine learning to rationalize and understand structure-property relations.
Supervised machine-learning techniques, that explicitly attempt to predict the properties of a material given its structure, are less susceptible to such biases. 
Current developments in the field suggest that using these two classes of approaches side-by-side and in a fully integrated mode, while keeping in mind the relations between the data analysis framework and the fundamental physical principles, will be key for realizing the full potential of machine learning to help understanding the behavior of complex molecules and materials.  \end{abstract}

\pacs{Valid PACS appear here}%
\maketitle

\section{Introduction}

Atomic-scale modelling of matter, including molecules, materials, and biological systems has progressed tremendously over the past decade, owing to the constant increase of available computational resources, the development of electronic-structure methods and empirical force fields with better accuracy-to-cost ratios, and the implementation of efficient, highly-parallel codes. 
A few recent landmark achievements include the use of dedicated hardware to obtain millisecond-long trajectories of biomolecules~\cite{shaw+09proc}, the discovery of exotic phases of matter by enumeration of locally stable polymorphs~\cite{pick-need11jpcm}, the compilation of high-throughput databases of hypothetical materials~\cite{jain+13aplm,moun+18nn}, and the simulation of crystal plasticity with models involving billions of atoms~\cite{shib+17nc}.

Over the past decade, a diverse array of methods that can be grouped under the loosely defined label of ``machine learning'' (ML) have been applied and adapted to atomistic simulations of materials and molecules. 
Given a set of structures $\left\{\CA_i\right\}$ (inputs) and, possibly, associated properties $\left\{y_i\right\}$ (labels), these algorithms aim to perform different classes of tasks in a way that relies as much as possible on the data, and minimizes human intervention and prior bias. 
One can roughly classify ML schemes into \emph{unsupervised learning} algorithms which only make use of the input points, and try to identify the organizing principles that underlie the data set and \emph{supervised learning} schemes which also use the labels and make predictions on the relations between inputs and some target properties. 
Unsupervised algorithms can be further subdivided into \emph{clustering} methods, that identify inputs that are closely related to each other; \emph{dimensionality reduction} schemes that reduce the dimension of the feature vectors that represent the inputs; \emph{generative models} that seek to construct new structures $\CA$ that are somehow compatible with $\left\{\CA_i\right\}$, or that are optimal in terms of parameter space exploration.
Supervised schemes include \emph{classifiers} that seek to partition the inputs based on their labels; \emph{regressors} that try to predict the value of the property $y$ for a new input $\CA$; \emph{inverse design} methods that try to generate new structures that exhibit a specified or optimal value of the property $y$.
Many of these tasks have been an integral part of the atomic-scale modelling workflow since the early days, and it is rather clear that it is mainly the data-centric, inductive mindset -- as well as a certain degree of hype -- that sets recent ML schemes apart. 
Put simply, the idea behind all these ML schemes is that, confronted with a sufficient quantity of high-quality data, an algorithm might be better at identifying the essential features of the problem than humans -- who are pretty good at looking for patterns and correlations, but who often fall prey to biases and logical fallacies~\cite{dawkins2016god}.

\begin{figure*}[tbhp]
    \centering
    \includegraphics[width=0.9\textwidth]{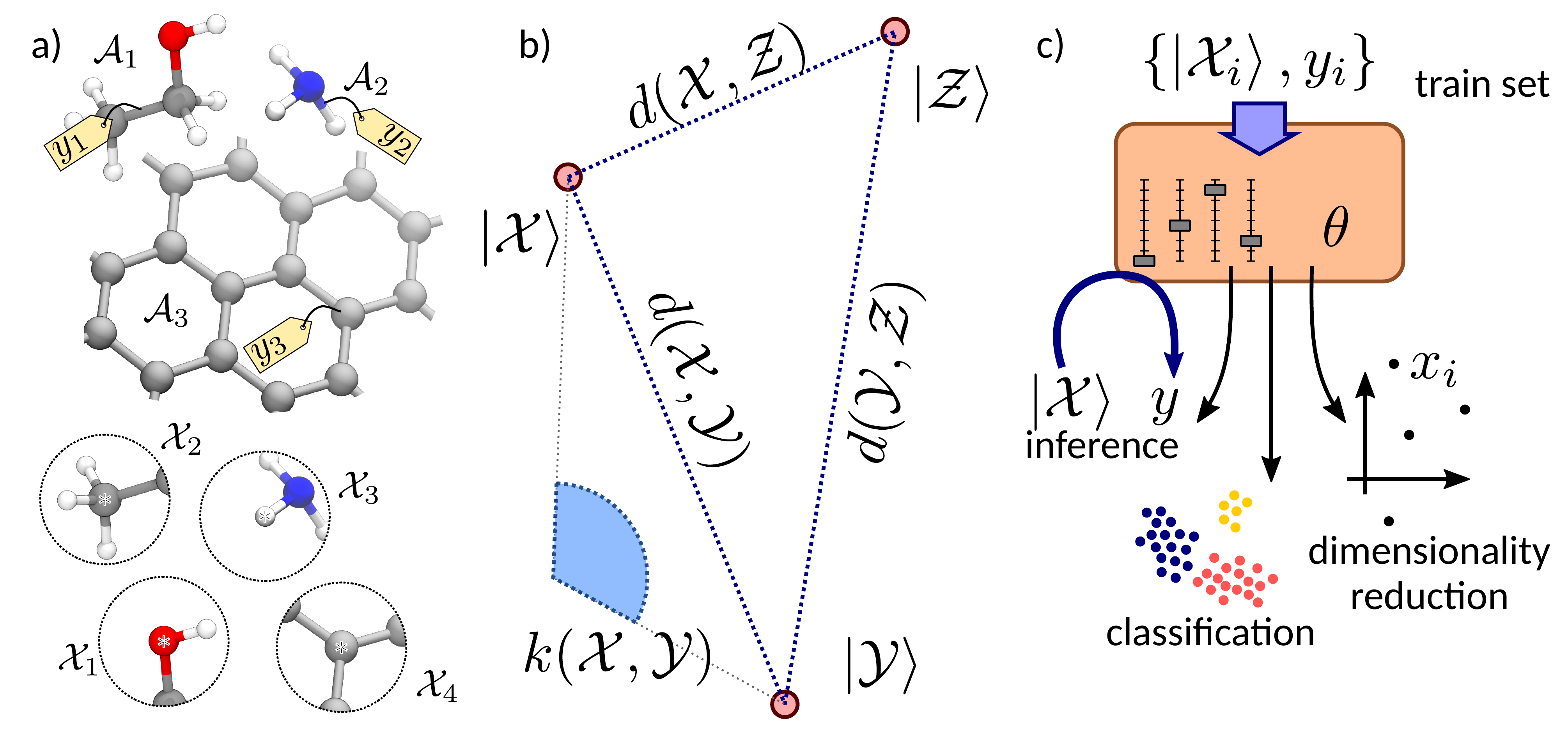}
    \caption{A schematic representation of the main ingredients of atomistic machine learning. (a) Structures $\CA$ or local environments $\CX$ are the inputs of the model, possibly with labels $y$ associated to them. (b) The inputs are associated with a mathematical representation, in terms of vectors of features $\ket{\CX}$, a measure of similarity $d$, or a kernel $k$. (c) The machine-learning model, controlled by a series of hyperparameters $\theta$, is trained based on a set of inputs. It can be then used for  a number of machine-learning tasks.}
    \label{fig:ml-overview}
\end{figure*}

Many books provide a primer in the field of machine learning~\cite{bish11book,rasm06book}, and several excellent reviews have appeared that have covered some recent applications of these schemes to atomic-scale modelling~\cite{rohr+13arpc,noe-clem17cosb,vonl18ac,butl+18nature}, including a perspective article in this same Journal that discusses the use of supervised learning to construct interatomic potentials~\cite{behl16jcp}. 
This perspective focuses on unsupervised-learning schemes, applied to the understanding of the behavior of complex materials and molecules. As  will become clear, my opinion is that the lines separating different families of ML approaches are blurred, and that it is highly beneficial to combine supervised and unsupervised tasks, as the former provide a quantitative framework to benchmark and validate the latter. 
This perspective is organized as follows. First, I will give a brief pedagogic introduction to machine learning, focusing on the ingredients that are central to its use in the context of atomistic modeling. Then, I will review some of the recent developments and applications of unsupervised learning in the field. Finally, I will articulate my opinions on  the main current challenges, and the opportunities that lie ahead. 

\section{Machine learning for atomistic modeling}

The typical workflow for the application of a ML scheme to an atomic-scale system is summarized in Fig.~\ref{fig:ml-overview}. 
The input can be a structure $\CA$ (a molecule, or a supercell describing a portion of a bulk structure) or a portion $\CX$ of a structure, e.g. a group of chemically-bound atoms, or a spherical, atom-centered environment. 
Each structure or group of atoms can have one or more properties $y$ associated with them, which serve as labels in supervised learning schemes. 

Even though atomic coordinates provide a complete description of each structure or environment, they are hardly ever used as the input of a ML scheme. Instead, they are usually converted into an appropriate \emph{representation}, which is more suitable for the task at hand. 
A first reason why an intermediate description of the system can be advantageous is that it makes it possible to incorporate obvious physical invariances (rigid translations and rotations, permutations of identical atoms) so that configurations that are identical are recognized as such. 
A second reason -- that is particularly relevant for unsupervised-learning applications geared towards simplifying the description of complex systems -- is that the choice of representation provides ample leeway to incorporate prior knowledge or to otherwise restrict the scope of the analysis. For instance, one may want to focus solely on the backbone of a protein, or on reactants in solution. Obviously, opting for a specialized description of a system introduces some biases in the procedure. However, in many cases these restrictions are either straightforward, necessary to make the analysis meaningful, or originate from technical limitations of the input data.

It is worth stressing that the representation of a structure can take the form of a vector of features $\ket{\CX}$ (e.g. all the backbone dihedrals in a protein, the coordination number of atoms in a cluster, etc.), a measure of similarity between pairs of structures that can be implemented as a distance $d(\CX,\CX')$, or the form of a kernel $k(\CX,\CX')$. 
In fact, these three ways of representing a data set can be used interchangeably (Fig.~\ref{fig:ml-overview}b). Given a representation in terms of a finite set of features $x_j(\CX) = \bra{j}\ket{\CX}$, it is always possible to use it as the starting point to define a distance or a kernel. 
While many different choices are possible -- some of which will be discussed in what follows -- one can consider as a simple example the case of the Euclidean distance $d(\CX,\CX')^2=\sum_j (x_j-x_j')^2$ and of a scalar-product kernel $k(\CX,\CX')=\sum_j x_j x_j'$. 

Less obvious is the fact that given a metric or a well-behaved (positive-definite) kernel it is possible to obtain an operative definition of a representation of items in the set in terms of a set of features  $\bra{j}\ket{\CX}$. 
From a formal point of view, the representer theorem guarantees that each positive-definite kernel~\cite{cutu10arxiv} is associated with a Hilbert space whose scalar product $\bra{\CX}\ket{\CX'}$ equals the kernel, and it is always possible to convert a kernel into a metric, e.g. by the relation $d(\CX,\CX')^2=k(\CX,\CX)+k(\CX',\CX')-2 k(\CX,\CX')$.
To see how in practice one can build a feature vector $\ket{\CX}$ given a kernel, consider a set of inputs $\mathbb{X}=\left\{\CX_i\right\}$. 
The matrix containing the value of the kernel between each pairs of inputs $K_{ii'}=k(\CX_i,\CX_{i'})$ is positive-definite, and can be diagonalized as $\mathbf{K} \mathbf{u}^{(j)} = \lambda_j \mathbf{u}^{(j)}$. 
Now take 
\begin{equation}
x_j(\CX) = \sum_k k(\CX,\CX_k) u^{(j)}_k /\sqrt{\lambda_j}. \label{eq:kpca}
\end{equation}
For each pair of inputs 
\begin{equation}
\begin{split}
{\bx}(\CX_i)\cdot{\bx}(\CX_{i'})=&
\sum_{jkk'} \lambda_j^{-1} K_{ik} u^{(j)}_k
K_{i'k'} u^{(j)}_{k'}\\
=&\sum_j \lambda_j u^{(j)}_{i} u^{(j)}_{i'} = k(\CX_i,\CX_{i'}).
\end{split}
\end{equation}
In other terms Eq.~\eqref{eq:kpca} reproduces exactly the value of the kernel for each structure in the reference set, and provides a practical approximation of the vector $\ket{\CX}$ associated with the kernel for any other structure. While this construction is not always practical (e.g. because it is not possible to truncate the expansion to a small number of eigenvectors) Eqn.~\eqref{eq:kpca} embodies the core idea behind the kernel trick: the (generally non-linear) kernel provides, together with a set of representative points $\{\CX_k\}$, a linear basis to describe new structures in terms of their relation to those chosen as reference. 

Once atomic configurations have been converted to a suitable representation, they can be fed to an appropriate machine-learning algorithm (Fig.~\ref{fig:ml-overview}c). 
Consider for instance the case of a long molecular dynamics simulations during which a molecule undergoes several conformational transitions. 
Unsupervised learning algorithms may use only the set of input representations of each structure $I=\left\{\ket{\CX_i}\right\}$ to identify structural motifs. 
This can be achieved by grouping structures that are similar to each other (e.g. configurations corresponding to thermal fluctuations around one of the metastable conformers) into disjoint clusters $C_j\subseteq I$ (clustering), and/or by obtaining a low-dimensional representation of the input vectors $\ket{\CX_i}\rightarrow \bx_i$ (dimensionality reduction), so that one can visualize the relations between minima and the transition pathways between them, much like with a geographic map. 
Most unsupervised-learning schemes allow one to perform out-of-sample operations, i.e. to take a new input $\ket{\CX}$ associated with a new configuration, and assign it to one of the clusters or map it to its low-dimensional representation $\bx(\CX)$.
Even though this perspective focuses on these families of techniques, it will also involve some discussion of supervised schemes. 
These include classifiers -- which determine the assignment of a new structure  $\ket{\CX}$ to one between two or more classes (e.g. the different conformers) after a training phase that relies on prior knowledge of the assignment of a subset of the inputs to the clusters $C_j\subseteq I$, as well as regressors, which provide a prediction of the properties $y(\CX)$ based on knowledge of the property values $y_i$ for each structure in the training set. 

Both supervised and unsupervised algorithms can  generally be tuned by adjusting several so-called hyperparameters  $\theta$, in addition to those that enter the definition of the input representation. 
Optimization of these hyperparameters is often a separate task from that of training the machine-learning scheme on a given input set. In the context of unsupervised learning it is often performed manually, by trial and error, and constitutes one of the critical points to consider to keep the procedure truly unbiased. 

\section{Representations, distances and kernels}

The problem of obtaining a complete yet concise description of a molecular or condensed-phase structure is central to many modelling techniques, including the sampling of rare events~\cite{laio-parr02pnas} and the search for stable structures~\cite{wale03book,ogan-glas06jcp,pick-need11jpcm}. 
A multitude of problem-specific representations have been proposed, including Steinhardt parameters for the description of crystalline order~\cite{stei+83prb}, cubic harmonics to differentiate different phases of solids and liquids~\cite{angi+10prb}, measures of chirality~\cite{piet+11jcc} and secondary structure~\cite{piet-laio09jctc} in proteins, and many others. 
In the context of this perspective, I would like, however, to focus on more abstract, general-purpose approaches that can be applied to molecules, solids and liquids alike. Many such representations have been introduced over the past decade, driven by a growing need to analize large atomstic datasets, and also by the development of supervised-learning approaches such as fitting of potentials.

\begin{figure}[tbp]
    \centering
    \includegraphics[width=0.85\columnwidth]{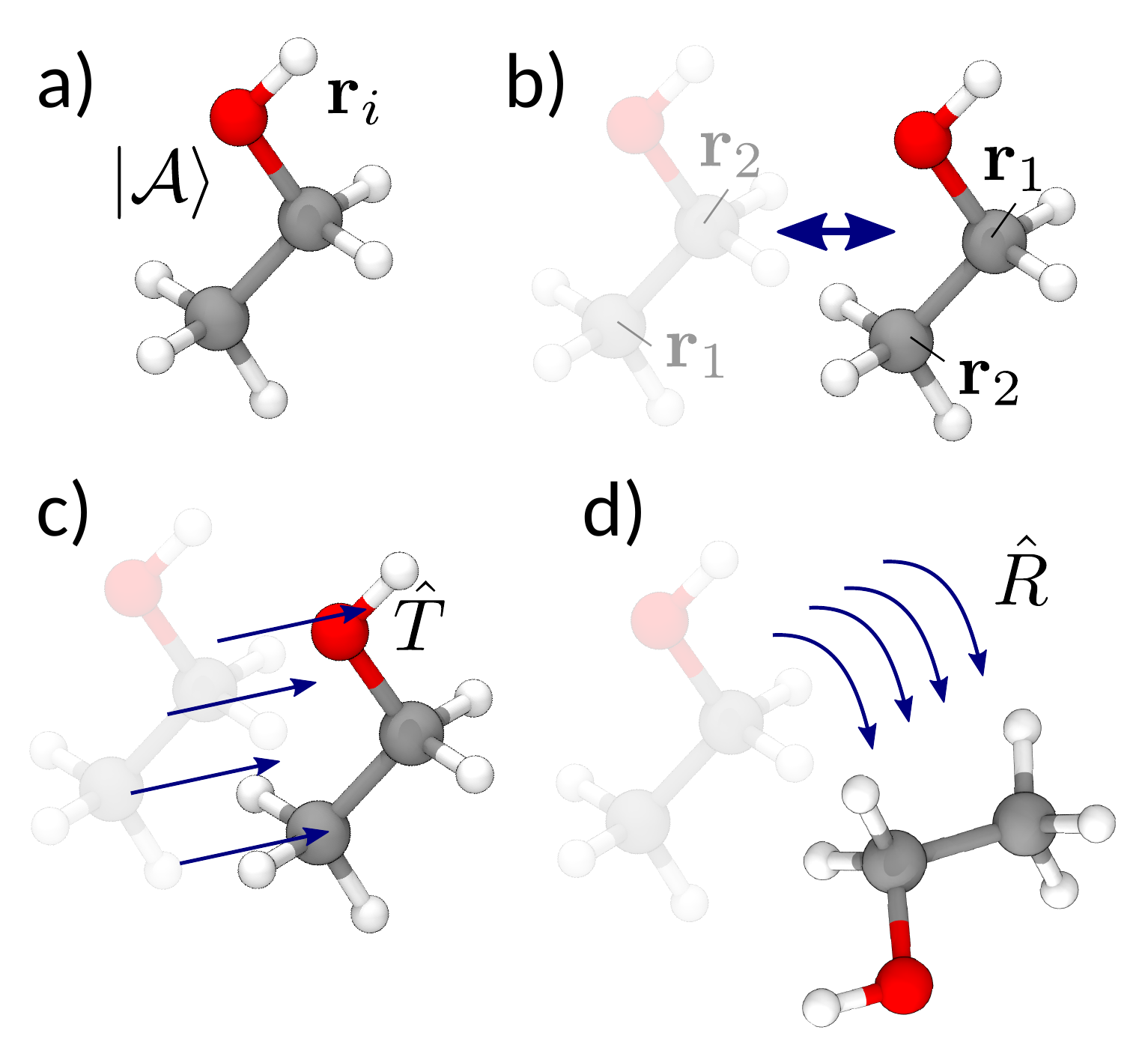}
    \caption{Fundamental symmetries that should be encoded in an atomistic representation. (a) Consider a structure $\CA$, described by the atomic types and their position $\br_i$. The associated feature vector $\ket{\CA}$ should be invariant to (b) permutation of the indices of same-kind atoms; (c) rigid translations and (d) rigid rotations.
    }\label{fig:symmetry}
\end{figure}

A first distinction between general-purpose representation schemes can be made separating between those that aim to describe a structure in its entirety, and those that instead focus on the description of fragments or local environments. 
\emph{Global} representations and metrics include those based on matrices of interatomic distances~\cite{piet-andr11prl,piet-mart15jcp}, transformed to yield Coulomb matrices~\cite{rupp+12prl} or Gaussian overlap matrices~\cite{sade+13jcp}, as well as pair correlation functions~\cite{schu+14prb,fabe+15ijqc} and extensions such as the many body tensor representation~\cite{2017Haoyanarxiv:1704.06439}. 
\emph{Local} representations, on the other hand, are geared towards providing a description of a structure in terms of atom-centered environments, and include Behler-Parrinello symmetry functions and their close relatives~\cite{behl-parr07prl,smit+17cs}, and those based on the bispectrum or the power spectrum of the atom density (smooth overlap of atomic positions kernels, SOAP)~\cite{bart+13prb}. 
Obviously, the separation is far from clear-cut, and it is often possible to construct a local version of a global representation~\cite{sade+13jcp}, and vice versa to assemble local kernels and feature vectors into a representation of entire structures~\cite{de+16pccp}. 

Another central consideration that one has to make when comparing representation involves whether and how they incorporate some of the fundamental physical symmetries: permutation of identical atoms, rigid translations and rotations (Fig.~\ref{fig:symmetry}). Other symmetries such as reflections or point/space-group operations might also be useful for some specific application.
The main downside with using raw atomic coordinates as an input representation is precisely the fact that they discriminate between structures that are physically indistinguishable. 
One may argue that given sufficient amounts of training data a good ML scheme should be able to recognize these invariances, but why bother: most of the state-of-the-art representations incorporate symmetries and by using them you can let your 20-layer convolutional neural network focus on learning correlations that are not as obvious. 

When comparing existing representations from the point of view of symmetries, one can recognize another binary divide between different approaches. 
One class of representations starts from a description of structures or environments in terms of internal coordinates (typically pair distances), that are naturally invariant to rotations and translations. 
Permutation symmetry is then included, either by summing over all possible permutations as in the case of permutation-invariant polynomials~\cite{braa-bowm09irpc} or by sorting the distances, or the eigenvalues of some distance-derived matrix~\cite{piet-andr11prl,piet-mart15jcp,sade+13jcp,zhu+16jcp}. 
A second class of representations, based on atomic densities, takes as a starting point a representation of structures in terms of atom-centered functions (generally Gaussians, or their $\delta$-distribution limit). The resulting atomic density is naturally invariant with respect to atom index permutations, but is neither translationally nor rotationally invariant. 
These symmetries are thus included by either projecting the atom density onto symmetry functions that only depend on interatomic distances and angles~\cite{behl11pccp}, or by explicit symmetrization of a kernel built on top of the atomic density~\cite{bart+13prb,glie+18prb}.

While the cost and effectiveness of different representations may vary, one can demonstrate that most of these schemes can be seen as a specific implementation of a very abstract density-based scheme, in which a ket $\ket{\CA}$ is associated to each configuration~\cite{2018Michaelarxiv:1807.00408},
\begin{equation}
\bra{\br}\ket{\CA} = \sum_{i\in \CA}
g(\br - \br_i)\ket{\alpha_i},
\end{equation}
where $g(\br)$ is a function peaked at atomic positions (e.g. a Gaussian) and $\ket{\alpha_i}$ is a vector that describes the chemical nature of each atom. 
Symmetries can be incorporated in this framework by averaging tensor products of $\ket{\CA}$ over the continuous translation and rotation groups, thereby obtaining symmetrized vectors that correspond to different $n$-body correlation functions between the atoms in each environment~\cite{2018Michaelarxiv:1807.00408}.
It is also possible to determine a direct relation between representations based on symmetrized atomic density and sorted internal coordinates, at least in simpler cases such as the approaches that rely on a vector of sorted interatomic distances.
Such a unifying view of input representations helps to understand the limits and strengths of each particular representation -- for instance how SOAP power spectra components are systematic but overdetermined, and how Behler-Parrinello symmetry functions are highly correlated with each other. 
Symmetries provide a solid physical basis on which representations of atomic structures that are at the same time complete and concise can be built.

\section{Clustering and pattern recognition}

\begin{figure}[b]
    \centering
    \includegraphics[width=1.0\columnwidth]{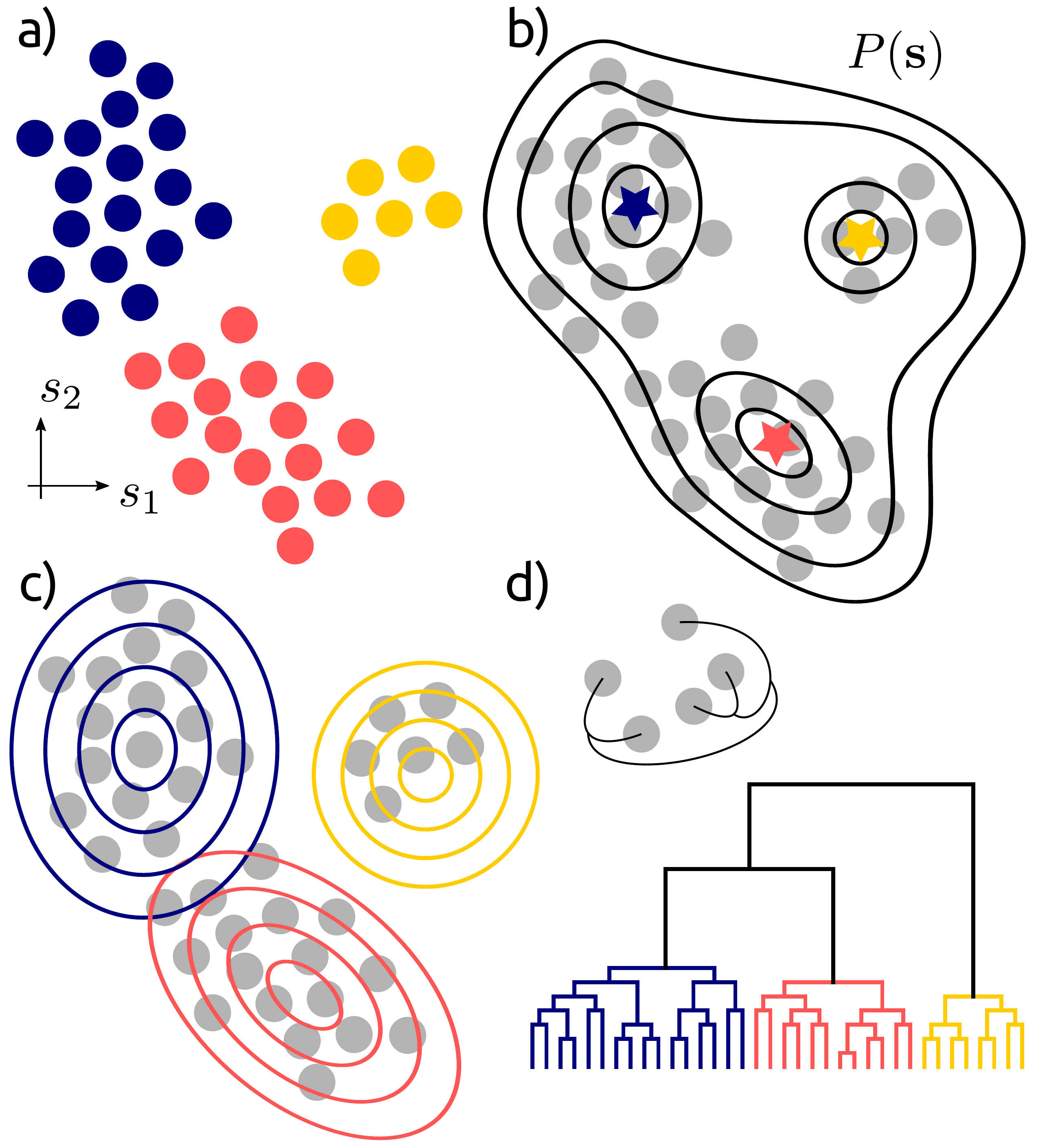}
    \caption{A summary of clustering techniques. (a) A set of points in a finite-dimensional feature space are clustered together in a way that reflects some underlying common characteristic. (b) Density-based clustering identifies maxima in the probability distribution of inputs in feature space. (c) Distribution-based clustering determines a model of the data distribution as a combination of cluster probabilities. (d) (Hierarchical) linkage clustering determines works by accretion of clusters starting on inputs that are closest together in input space. }
    \label{fig:clustering}
\end{figure}

Clustering algorithms aim to recognize groups of input points that are related to each other, and different from other groups of inputs. For instance, clusters could represent different classes of molecules, or configurations of a given system that are separated  by a sparsely-populated, or seldom accessed region (see Fig.~\ref{fig:clustering}a). 
A common mental image to understand materials and molecules that can undergo non-trivial dynamics involves a rugged landscape in which valleys correspond to (meta)stable states, and saddle points to the most energetically-favorable pathways connecting these states~\cite{wale03book}. 
In other terms, given a representation in terms of a set of features $\mathbf{x}$ that describes a chemical reaction, a conformational change or a phase transition, the free energy $F(\mathbf{x})$ comprises local minima, around which the system can fluctuate at finite temperature, and transition states across which the system can occasionally undergo jumps to a nearby state. 
The stability of the various configurations is reflected in the finite-temperature probability distribution $e^{-\beta F(\mathbf{x})}$ that exhibits separate peaks for each minimum. 

The connection between density peaks and minima in the free energy makes density-based clustering methods (Fig.~\ref{fig:clustering}b) particularly well-suited to the analysis of finite-temperature atomistic simulations. Taking for simplicity the case of a finite-dimensional representation $\bx = \bra{\bx}\ket{\CA} \in \RR^D$, the clustering algorithm has to identify the local maxima in the probability distribution $P(\bx)$, which is inferred from the collection of input points $I=\{\bx_i\}$. 
The problem can be broken down into that of estimating the probability distribution, and that of recognizing the maxima. 
For example, the ``fast search and find of density peaks'' method~\cite{rodr-laio14science} defines the density at a point $\bx$ based on the number of inputs in $I$ that fall within a small distance $\epsilon$ of it. 
The ``probabilistic analysis of molecular motifs'' (PAMM) method~\cite{gasp-ceri14jcp,gasp+18jctc} performs a kernel density estimation on a subset of the input points that are taken as representative landmarks $L=\{\bx_j\}$, i.e. it computes $P_j=\sum_{\bx\in I} g(\bx_j -\bx)$, where $g$ is a Gaussian, or another localized function. 
Then, both algorithms apply a mode-seeking algorithm, commonly known as quick-shift,~\cite{carr00ieee} to identify local maxima in the distribution, defined as those points $j$ for which there is no other point $j'$ within a threshold distance $\lambda$ for which $P_{j'}>P_j$. 

An alternative approach to identify clusters relies on making assumptions on the form of the underlying distribution, e.g. taking $P(\bx)=\sum_i p(i) p\left(\bx\middle|i\right)$, where $p(i)$ corresponds to the fraction of probability associated with the $i$-th cluster, and $p\left(\bx\middle|i\right)$ is the distribution of values of $\bx$ that is assumed for members of the cluster.
Gaussian-mixture models (GMM) take the cluster probability distribution $p\left(\bx\middle|i\right)$ to be a multi-dimensional Gaussian (Fig.~\ref{fig:clustering}c). Its parameters (mean and covariance) are determined, together with the cluster populations $p(i)$, by expectation maximization on the input set~\cite{carr00ieee}.
An appealing feature of distribution-based clustering algorithms is that they provide a probabilistic model to assign new points to existing clusters,  $\hat{p}\left(i\middle|\bx\right) = p(i) p\left(\bx\middle|i\right) / \sum_j p(j) p\left(\bx\middle|j\right)$.
GMMs, and similar parametric cluster models are often criticised as the number of clusters and the initial Gaussian parameters affect the outcome of the analysis. This problem can be mitigated by using the outcome of a non-parametric clustering technique as the starting point for a GMM~\cite{gasp-ceri14jcp,gasp+18jctc}.

One of the simplest clustering methods, $k$-means clustering~\cite{macq67proc}, can be seen as a special case of a GMM, in which the Gaussians are all taken to have the same diagonal covariance matrix, and a spread that is small compared to the distance between cluster centers. 
Another commonly-used density-based clustering technique is DBSCAN~\cite{este+96proc}, in which regions with high density are identified by checking whether many points fall within a prescribed distance of a given point, and contiguous regions with high density define separate clusters. Inputs that do not belong to any of these high-density regions are discarded as noise. 

DBSCAN combines ideas from density-based clustering and hierarchical clustering methods, a family of algorithms that rely heavily on growing existing clusters starting from a core of closely-connected points. 
Agglomerative clustering~\cite{murt-cont17wir} proceeds by finding the inputs that are closest to each other in feature space, and grouping them together. 
Using different definitions for the distance between groups of inputs, it is then possible to combine further such initial clusters, to form a tree-like structure that represents the relations between clusters as well as the classification of the inputs (Fig.~\ref{fig:clustering}d).
Rather than growing clusters iteratively, it is also possible to define a connectivity matrix between the inputs and to use graph theoretical considerations to identify groups of points that are closely related to each other, and/or weakly linked to other groups of inputs~\cite{cafl06cosb,chen+15pnas}.

In closing this brief overview, it should be stressed that all of these techniques depend crucially on the choice of the feature space, and/or the metric that is used to estimate the similarity between inputs. Non-linear transformations of the input space, or of the measure of similarity, introduce distortions that can easily split, merge, create and destroy local maxima in the probability distribution associated with the input set. As we will discuss below, the risk of introducing spurious features can be mitigated by using general-purpose, relatively simple functions of the atomic coordinates as the input representation, and by checking that clustering is insensitive to the choice feature space, kernel or metric.

\section{Applications of clustering methods to atomistic modeling}

Clustering algorithms have been applied since the early days of molecular simulations to recognize (meta)stable configurations of complex molecules and biomolecules~\cite{karp+93bc,tord-vang94jcc},  to coarse-grain large-scale trajectories into a set of discrete states, and in very broad terms to identify the most relevant configurations out of long trajectories.
This discretization of configuration space is also often used as the basis to map the continuous dynamics of a molecular system into a sequence of jumps, and to thereby reduce the complexity of correlated molecular motion to transition rates between free energy basins. 
The development of such Markov-state models~\cite{pand+10methods,chod-noe14cosb} (MSM) is closely interlinked with that of clustering algorithms, because an accurate and unbiased determination of kinetically distinct states is crucial when it comes to guaranteeing that the underlying assumption of a Markovian dynamics holds true, particularly in the case of biomolecular simulations in which entropically-stabilized states are the norm rather than the exception. 
Kinetic Monte Carlo models~\cite{vote+02armr} -- essentially the materials science version of a MSM -- are instead usually built based on potential energy landscape exploration and on rates derived from transition-state theory. 
This is because in crystalline materials with few defects local minima on the potential energy surface  unequivocally identify the meta-stable states of the system. 
It is likely that in the near future there will also be a growing need for automated structural analysis also in this field, as materials with complex, correlated transitions become amenable to atomistic modeling~\cite{kahl+18prm}.

Clustering of structurally-related configurations is not only useful to obtain dynamic coarse-grained models. Automated structure searches~\cite{goed04jcp,ogan-glas06jcp,pick-need11jpcm}, and high-throughput materials modeling~\cite{curt+13nat,phil16aplm,Pizzi2016} generate vast numbers of configurations. Determining which structures are genuinely distinct, and which are only small, inconsequential variations on the same skeleton is a crucial and time-consuming task associated with computational searches. 
For these reasons, clustering and classification schemes have been extensively used in this context,  to analyze existing databases~\cite{isay+15cm,de+16jci}, to post-process trajectories in order to identify polymorphs~\cite{piet-mart15jcp,piag-parr18pnas}, as well as to discard duplicates ``on the fly'' and to terminate early the optimization of redundant configurations~\cite{goed04jcp,amsl-goed10jcp,curt+18jctc}.
Machine-learning representations that are based on the description of local motifs, rather than on a conventional characterization of the system in terms of its lattice parameters and minimal unit cell, are very convenient, because they simplify the task of recognizing structures that correspond to alternative descriptions of the same periodic configuration (e.g. a conventional versus a primitive \emph{fcc} unit cell). 

The use of local, atom/fragment-based representations is also convenient to find recurring patterns in the building blocks of a structure, rather than in the overall configuration. 
This kind of analysis has a long-standing tradition in the context of (bio)macromolecules, e.g. to detect and classify binding pockets in proteins~\cite{rupp+08ps} or recurring motifs in RNA base pairs~\cite{yang03nar}. 
It has the potential of recognizing the molecular motifs that underlie the behavior of entire classes of materials~\cite{huan+15prb}, and has been used to rediscover, extend or redefine altogether entities such as the hydrogen bond~\cite{gasp-ceri14jcp} or secondary-structure motifs~\cite{gasp+18jctc} that are at the heart of physical/chemical intuition and of well-established structure-property relations.  
In materials local pattern recognition schemes such as these have been used to identify crystal packing motifs~\cite{rein-pana18sm}, as well as point~\cite{zimm+17fm} and line~\cite{stuk+12msmse} defects, and once again they will become increasingly important as high-throughput/large scale simulations of disordered or otherwise complex materials become commonplace.

\section{Dimensionality reduction}

Clustering techniques identify agglomeration of data in the most significant regions of feature space, but do not offer an overall picture of the relations between different structures. Such relations can be determined, and represented, as edges connecting clusters that are structurally or kinetically adjacent to each other.
In principle, the information on the absolute position of each input in space, and the relations between any given pair of inputs, are encoded fully in the feature vectors. It is however hard if not impossible to comprehend -- not to mention visualize -- the meaning or the relative positions of vectors in a high (or infinite) dimensional space. 
This is a problem that obviously transcends the specific application to chemistry, physics or materials science, and dimensionality-reduction methods have long been sought after for all sorts of data analysis applications.

\begin{figure}[tbp]
    \centering
    \includegraphics[width=1.0\columnwidth]{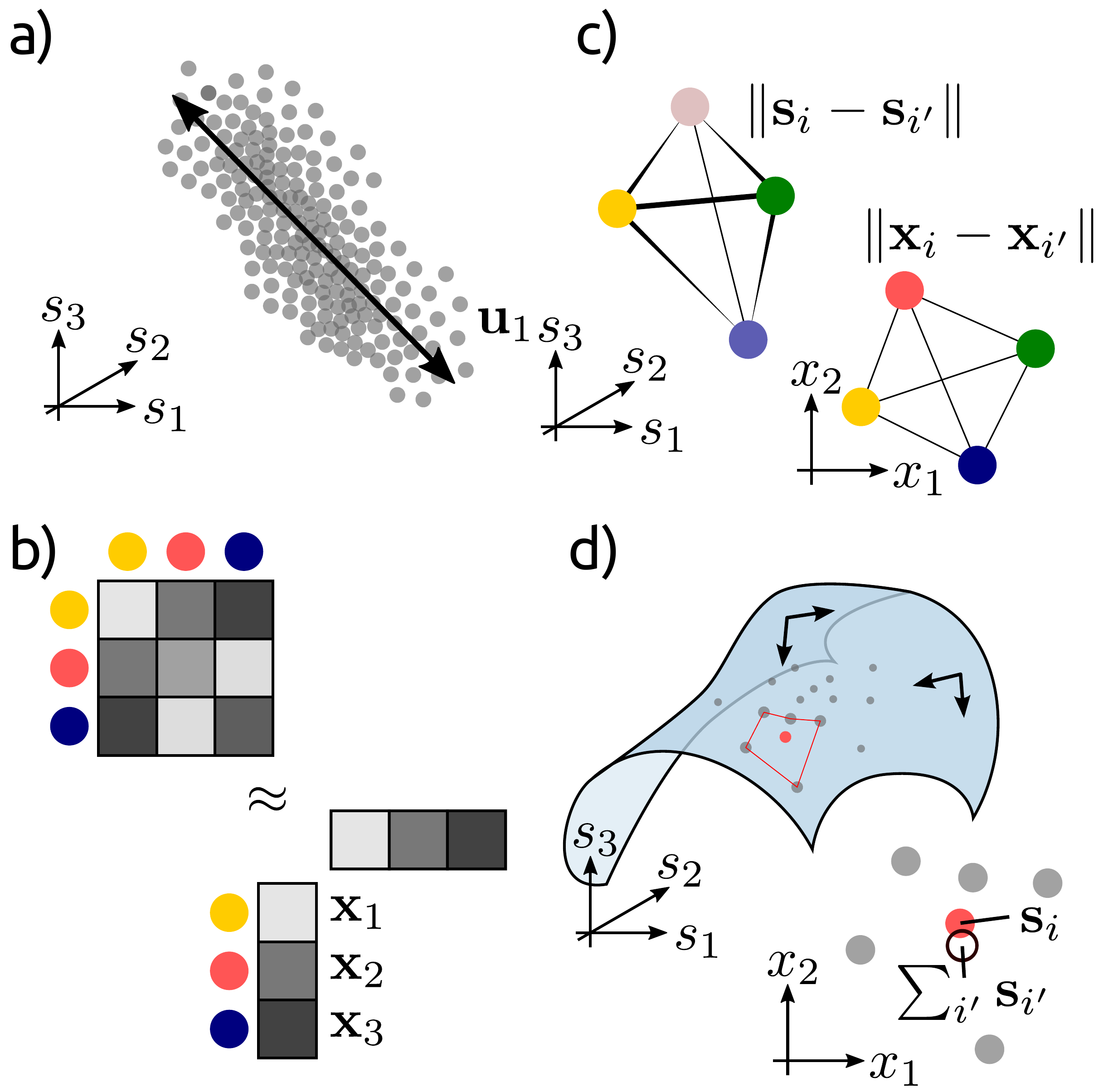}
    \caption{A summary of the main strategies underlying dimensionality reduction techniques. (a) Several methods (starting from PCA) attempt to identify the low-dimensional subspace that captures the largest fraction of the input data variance. (b) This is equivalent to finding the best $\mathcal{L}^2$-norm approximation of the Gram matrix, and generalizes to kernel methods. (c) Multi-dimensional scaling and related approaches attempt to reproduce the similarity between high-dimensional data points in low dimension. (d) Embedding methods explicitly try to preserve local relations between points, under the assumption that they lie on a (locally) low-dimensional manifold. }
    \label{fig:dimred}
\end{figure}

Principal component analysis (PCA) is probably the simplest of all dimensionality-reduction schemes, and can be taken as a paradigmatic example to summarize some of the fundamental ideas that underpin other dimensionality reduction approaches. Given a set of $D$-dimensional input vectors $\bx_i$ associated with $N$ structures or local environments, PCA identifies the $d$ linear combinations of the feature components that capture the largest possible fraction of the variability of the dataset (Fig.~\ref{fig:dimred}a). 
The problem boils down to computing the covariance between the features, $C_{jj'} = N^{-1} \sum_i x_{ij} x_{ij'}$, and selecting the eigenvectors $\bu_k$ associated with the  $d$ largest eigenvalues. The  principal-component representation of the input data is given by $s_{ik} = \bx_i \cdot \bu_k $.

PCA lends itself to alternative formulations, making it possible to link it to several classes of non-linear dimensionality reduction techniques. It can be seen as a problem of matrix approximation: the matrix of the scalar products between principal component vectors $\bs_i\cdot\bs_{i'}$ is the best rank-$d$ approximation of the Gram matrix of input feature vectors $G_{ii'}=\bx_i\cdot\bx_{i'}$ (Fig.~\ref{fig:dimred}b). 
In this form, there is a transparent connection between PCA and the so-called kernel PCA (KPCA)~\cite{scho+98nc} method, that considers the principal values of the kernel matrix between the inputs rather than the Gram matrix, and therefore incorporates nonlinearities through the definition of the kernel~\footnote{Kernel PCA corresponds essentially to a truncation of Eq.~\eqref{eq:kpca} to the largest $d$ eigenvalues.}.
Another way to interpret PCA is to see $\left\{\bx_i\right\}$ as the set of $d$-dimensional points that provide the best approximation of the similarity matrix $S_{ii'}=\left\|\bx_i-\bx_{i'} \right\|^2$. In PCA  this matrix is based on a Euclidean distance and assumes $\bs_i$ to be a \emph{linear} projection of the corresponding $D$ dimensional point $\bx_i$.

If one relaxes the assumption that $\bs_i$ is a linear projection of $\bx_i$, the problem can be cast as a  optimization of a loss function
\begin{equation}
\ell^2 = \sum_{ij} \left( \left\|\bx_i - \bx_j\right\|^2 - \left\|\bs_i - \bs_j\right\|^2   \right)^2.
\end{equation}
This  corresponds to the simplest form of multi-dimensional scaling (MDS)~\cite{cox-cox10book}, which is the basis of several kinds of non-linear dimensionality reduction algorithms (Fig.~\ref{fig:dimred}c).  Sketch-map~\cite{ceri+11pnas,trib+12pnas}, for instance, applies non-linear transformations to the distances in high and low-dimension
\begin{equation}
\ell^2 = \sum_{ij} \left[ F\left(\left\|\bx_i - \bx_j\right\|\right) - f\left(\left\|\bs_i - \bs_j\right\|\right)   \right]^2.
\end{equation}
$F$ and $f$ can be tuned to disregard short-range thermal fluctuations (which tend to be full-dimensional, if the feature space is non-degenerate) and focus the optimization of the loss on the intermediate range of distances that are usually the most relevant to describe the transitions between meta-stable states~\cite{ceri+13jctc}.
The t-distributed stochastic neighbor embedding (t-SNE)~\cite{maat-hint08jmlr} also focuses on proximity information, by defining in high dimension the probability that two points are neighbors (usually defined as a Gaussian function of the distance), and then trying to ensure that neighbor probabilities in low dimension are similar, using a Kullback-Leibler divergence to compare the distributions. 

Using a metric that is different from the Euclidean distance between feature vectors is another way of improving the performance of MDS. For instance, ISOMAP~\cite{tene+00science} uses a neighborhood analysis to estimate geodesic distances between points, based on the assumption that the data lies on a curved, but locally low-dimensional, manifold.
The idea that the data may lie on a manifold that is only low-dimensional on a local basis is also behind methods such as locally-linear embedding (LLE)~\cite{rowe-saul00science} that function by writing each input as a weighted sum of its neighbors
\begin{equation}
\tilde{\bx}_i \approx \sum_{\left\|\bx_j-\bx_i\right\|<\epsilon} w_{ij} \bx_j, 
\end{equation}
and then by imposing self-consistently that each point in low dimension should be a combination of its neighbors using the same weights (Fig.~\ref{fig:dimred}c)
\begin{equation}
\bs_i \approx \sum_{\left\|\bx_j-\bx_i\right\|<\epsilon} w_{ij} \bs_j. 
\end{equation}

A last class of dimensionality reduction schemes deserves a separate mention. All of the methods discussed this far rely exclusively on the distribution of atoms in each structure. It is often the case, however, that the most relevant degrees of freedom for a system are those in which the dynamics are slower. 
For this reason, methods such as diffusion maps~\cite{coif+05pnas,ferg+10pnas,rohr+11jcp,rohr+13arpc} try to link geometric data to the characteristic time scale of the system's motion. Diffusion maps  essentially correspond to a kernel PCA, but the construction of the kernel and the interpretation of the principal components attempt to draw a link with the slowest eigenmodes of the Fokker-Planck equation for the diffusion of the system in feature space. 
The time-lagged independent component analysis method (TICA)~\cite{molg-schu94prl,pere+13jcp} identifies linear combinations of features from a time-dependent trajectory that not only have the largest variance, but that also have the longest autocorrelation time -- the main drawback being the large amount of simulation data that is necessary to reliably estimate the autocorrelation function. 
A similar combination of structural and dynamical information is used in the spectral gap optimization of order parameters (SGOOP) method to single out the linear combination of input features that correspond to the direction that is most relevant for the long-time dynamics of the system being studied~\cite{tiwa-bern16pnas}.

\section{Applications of Dimensionality Reduction to Atomistic Modeling}

The applications of dimensionality reduction schemes to atomistic simulations reflect the dual function of a coarse-grained representation, as a mean to simplify the description of a system to understand it better, and to explore more effectively its conformational space. 
On one hand, if you need to explain the behavior of a complex system, a picture is worth a thousand words, and paper tends to be two-dimensional -- although it would be entertaining if scientific publishers would offer, for a fee, a pop-up book option for truly complicated graphics. 
On the other hand, dimensionality reduction lies at the heart of the vast majority of accelerated sampling methods, which try to beat the time scale problem of molecular simulations -- that is, the enormous gap between the time scale of atomic vibrations and that of any interesting structural or phase transition. 

The former application has become particularly popular with the rise of high-throughput simulations and the construction of large databases of structures and materials~\cite{ramp+17npjcm}. 
``Materials cartography'', as it has been referred to~\cite{isay+15cm}, strives to obtain a low-dimensional (typically 2D) representation of a set of materials, or of different phases of the same material, that reflects the ``structure-energy-property'' landscape~\cite{musi+18cs,yang+18cm} of that system.
Similar approaches have been applied to biomolecular simulations~\cite{reut-schn12jmgm}, where the  main application of dimensionality reduction has been in connection with accelerated sampling.
In order to observe a rare (activated) event without developing dedicated hardware~\cite{shaw+09proc} one needs to selectively increase the speed at which the pathways related to the transition are traversed~\cite{vote+02armr,bolh+02arpc,abra-tuck08jpcb}, to modify the probability distribution that is sampled during the simulation so that configurations that are close to the transition state occur more often~\cite{torr-vall99jcp,laio-parr02pnas}, or to generate adaptively ensembles of trajectories that cover the relevant portions of phase space without an explicit bias~\cite{pret-clem14pccp}.
Either way, the objective is to accelerate sampling while introducing a minimal disturbance to the natural behavior of the system~\cite{tiwa-parr13prl}, since otherwise it becomes very inefficient to re-calibrate the statistics and the dynamics so as to infer the behavior of the real -- non-accelerated -- system~\cite{ceri+12prsa}.
Using a high-dimensional space to accelerate sampling almost invariably leads to statistical inefficiency. This is essentially a manifestation of the curse of dimensionality. 
Take for instance the case of metadynamics, in which a repulsive bias is accumulated in locations that have already been visited during the trajectory, forcing the system out of free energy minima~\cite{laio+05jpcb,laio-gerv08rpp}. The volume of one of these minima grows exponentially with the dimensionality $d$ of collective variable space, and so does the time to accumulate sufficient bias to escape the free-energy basin. 
The challenge of identifying the most appropriate low-dimensional space to accelerate conformational sampling is reflected in the fact that most of early (and recent) applications of dimensionality reduction to atomistic modeling involved simulations of rare events or phase transitions~\cite{das+06pnas,ceri+11pnas,spiw-kral11jcp,zhen+11jpcb,trib+12pnas,chia+17pnas}.

\section{The problems with unsupervised learning}

The typical selling point of most applications of unsupervised learning to materials, molecules and in general atomistic modelling is that a ``data-driven'' identification of regular patterns in simulation data is more robust, and less prone to preconceived biases, than a manual analysis carried out by a human. 
I myself have often used this narrative, and written titles and abstracts including words such as ``agnostic'', ``unbiased'' and ``automatic''.
In all honesty, most of the time the analyses did require a significant amount of human intervention, first in the design of the details of the classification/dimensionality reduction algorithm, and later in the choice of the hyperparameters that define the structural representation and tune the functioning of the machine learning scheme. 
The core issues are that: (1) the most straightforward unsupervised learning schemes detect correlations between data points in feature space, and so depend crucially on what representation is used, as well as on how the pattern recognition scheme defines proximity and correlations; (2) when working with actual data rather than synthetic data there is no ``ground truth'' in terms of what partitioning of the data, or low-dimensional representation, is performing best.
It is then tempting to pick one own's arbitrary perception of what an ``insightful'' outcome of the analysis would be, and tune the representation and machine-learning scheme until the results reflect most clearly one's expectations.

This is not to say that unsupervised learning techniques cannot offer genuine insight: in Ref.~\citenum{musi+18cs} we began our  by looking for regularity in the stacking of pentacene molecules in different stable polymorphs. When considering the nitrogen-substituted analogues of pentacene, however, we found that the most clear-cut clustering was associated with in-plane H-bonding rather than with the stacking. 
As is the case for any scientific endeavor, approaching a problem with an open mind and a healthy dose of skepticism can help in obtaining a deeper, non-obvious understanding. In particular, one should be wary of using a data representation that is excessively fine-tuned to the problem at hand.
General-purpose and systematically-convergent descriptions of chemical environments, such as those that are typically used to fit potentials or predict molecular properties~\cite{bart+13prb,behl11pccp}, provide a starting point that is considerably less biased than descriptors based on complicated heuristics that are designed from the outset to identify e.g. a stacking order, or the coordination of a site. 
Even more importantly, one should try as much as possible to assess the quality of clustering, or of a low-dimensional representation, based on an objective metric rather than on how well it reproduces the expectations of an expert in the field. 

\section{Putting some supervision into unsupervised learning}

In my opinion, the best way to address the problem of judging objectively how well a given structural representation reflects the structure-property relations for a certain problem involves checking how well that representation performs in a supervised learning task aimed at predicting that property. 
In the field of biomolecular simulations, the order parameters obtained by trial-an-error, or by unsupervised learning, are often benchmarked in terms of how well they perform to drive an accelerated sampling algorithm, or on how well they perform in terms of an analysis of the committor~\cite{bolh+00pnas}. 
This effectively introduces a quantitative metric to compare one candidate against another. It is then possible to automate the parameter optimization, so as to maximise the similarity with the committor~\cite{ma-dinn05jpcb,pete-trou06jcp,2019Hendrikarxiv:1901.04595}.
Dimensionality reduction algorithms as TICA~\cite{molg-schu94prl,pere+13jcp} or SGOOP~\cite{tiwa-bern16pnas}, that incorporate information on the system's dynamics, implicitly ensure that the order parameter(s) perform well as descriptors of the slowest molecular motion.

More generally, checking how accurately one can machine-learn the property of interest (cohesive energy, response properties, conductivity, \ldots) based on the structural representation that is used as the input for an unsupervised learning scheme can be a very informative exercise. 
For starters, it gives an idea of the limiting accuracy that can be expected when inferring structure-property relations by application of unsupervised classification or coarse-graining schemes. 
In combination with an analysis of the learning curves~\cite{huan-vonl16jcp} (the accuracy of the model when making a prediction, as a function of the size of the training set) it is possible to understand whether the problem is data-limited, or if the representation does not contain enough information to reflect the correlations between structure and properties. 
A critical assessment of the model performance as a function of the hyperparameters makes it possible to gauge the range of interactions~\cite{bart+17sa}, or the locality of the physical phenomena underlying structure-property relations~\cite{gris+18prl,wilk+19pnas} for that system. 
The accuracy of the regression model also provides an objective metric to optimize the hyperparameters that define the representation~\cite{will+18pccp}, and even to determine the most effective non-linear function of features that reproduces the target properties~\cite{ouya+18prm}.

Choosing to describe the inputs using a feature vector that is not problem specific, and validating and optimizing it through a relevant supervised learning exercise, reduces substantially the amount of human intervention and the possibility of introducing preconceived biases when  applying clustering and dimensionality reduction schemes.
To an extent, combining supervised and unsupervised learning can be mutually beneficial. Clustering can be used to select the most relevant reference configurations from a database~\cite{bart-csan15ijqc}, or to partition it into regions on which specialized models can be trained.~\cite{Kranz_2018} 
A low-dimensional representation of a dataset can help when it comes to identifying problematic regions, and understanding how a given representation ``sees'' a set of structures~\cite{bart+17sa,de+16jci,nguy+18jcp}. Unsupervised learning strategies can also be used to reduce the number of features that are needed to describe a system~\cite{imba+18jcp}.
Ultimately, an effective approach might be to devise data analysis techniques that combine elements of supervised and unsupervised learning. For instance, we have recently developed a generalized convex hull (GCH) construction~\cite{anel+18prm} that uses (kernel) principal component analysis to determine the structural parameters that capture the highest degree of diversity in a collection of (meta)stable structures. These coordinates are then used to build a convex hull that also incorporates an estimate of the stability of each structure to identify those that -- by being structurally diverse and comparatively stable -- have a high potential for being synthesizable. 
Another example of the synergy between supervised and unsupervised learning tasks involves the use of regression techniques to reconstruct (high-dimensional) free-energy surfaces, which mitigates the problem of the curse of dimensionality when performing essentially a density estimation~\cite{stec+14jctc,mans-ferg15jcp,mone+16jctc,schn+17prl}.

There is little doubt that the use of machine-learning techniques in atomistic modelling will outlive the hype, as the complexity of simulations is increasing for materials, chemical and biomolecular applications. 
Statistical learning will contribute to the increase of complexity by making it possible to side-step time-consuming electronic structure calculations, and obtain accurate interatomic potentials that can be evaluated on large systems, and for long trajectories.
At the same time, automated data analyses will be necessary to make sense of the outputs of longer, larger and more complicated simulations, and single out regularities and correlations that might be far from obvious. 
The convergence between different classes of machine-learning algorithms is one of the most promising directions to achieve genuine progress in the field. Unsupervised schemes can help shed light on the functioning of supervised algorithms that are often opaque, and supervised schemes can help to formulate quantitative metrics to assess the performance of unsupervised learning. 

\begin{acknowledgments}
I would like to express my gratitude to several colleagues who shared insightful comments on early versions of this article, including Alessandro Laio, Cecilia Clementi, Mariana Rossi, Gareth Tribello and David Manolopoulos. 
Funding by the European Research Council under the European Union's Horizon 2020 research and innovation programme (grant agreement no. 677013-HBMAP) is gratefully acknowledged.

\end{acknowledgments}

\end{document}